\documentclass[12pt,preprint]{aastex}
\begin{document}
\title{EFFECT OF CENTRAL MASS CONCENTRATION ON THE FORMATION 
OF NUCLEAR SPIRALS IN BARRED GALAXIES}
\author{PARIJAT THAKUR\altaffilmark{1}, H.B. ANN\altaffilmark{2}, AND 
ING-GUEY JIANG\altaffilmark{1}}
\altaffiltext{1}{Department of Physics and Institute of Astronomy, 
National Tsing-Hua University, Hsin-Chu 30013, Taiwan; 
pthakur@phys.nthu.edu.tw, jiang@phys.nthu.edu.tw}
\altaffiltext{2}{Division of Science Education, Pusan National University,
 Busan 609-735, Korea; hbann@pusan.ac.kr}
\begin{abstract}
We have performed smoothed particle hydrodynamics (SPH) simulations 
to study the response of the central kiloparsec region of a gaseous disk 
to the imposition of nonaxisymmetric bar  potentials. 
The model galaxies are composed of the three axisymmetric 
components (halo, disk, and bulge) and a non-axisymmetric bar.
These components are assumed to be invariant in time in 
the frame corotating with the bar. The potential of spherical 
$\gamma$-models of Dehnen is adopted for the bulge component whose density 
varies as $r^{-\gamma}$ near the center and $r^{-4}$ at larger radii
and hence, possesses a central density core for $\gamma = 0$ and cusps 
for $\gamma > 0$.  Since the central mass concentration of 
the model galaxies increases with the cusp parameter $\gamma$, 
we have examined here the effect of the central mass concentration 
by varying the cusp parameter $\gamma$ on the mechanism responsible 
for the formation of the symmetric two-armed nuclear spirals 
in barred galaxies. Our simulations show that the symmetric two-armed 
nuclear spirals are formed by hydrodynamic spiral shocks driven 
by the gravitational torque of the bar for the  models 
with $\gamma = 0$ and $0.5$. On the other hand, 
the symmetric two-armed nuclear spirals in the models 
with $\gamma=1$ and $1.5$  are explained by gas density waves. 
Thus, we conclude that the mechanism responsible for the formation 
of the symmetric two-armed nuclear spirals in barred galaxies 
changes from the hydrodynamic shocks to the gas density waves  
when the central mass concentration increases from $\gamma = 0$ to $1.5$. 
\end{abstract}

\section{INTRODUCTION}

High resolution visible and near-infrared observations by 
the {\it Hubble Space Telescope (HST)} and ground-based telescopes 
with adaptive optics show nuclear spirals in the centers of
active galaxies \citep{reg99,mar99,pog02,mar03a,mar03b}, as well as 
in those of normal galaxies \citep{phi96,elm98a,lai99,lai01,car02,mar03a,mar03b}.
Nuclear spirals have a variety of morphologies such as grand-design spirals, 
one-armed spirals, and flocculent spirals \citep{mar03a}.
The grand-design nuclear spirals are symmetric two-armed spirals,
while the flocculent, or ``chaotic,'' nuclear spirals have 
multiple arms that are likely to be segmented into smaller arms. 

The morphology of symmetric two-armed nuclear spirals depends on 
the gas sound speed and the shape of the gravitational potential 
in the nuclear regions of galaxies \citep{eng00,at05}. A number of hydrodynamic 
simulations \citep{eng97,pat00,eng00,mac02,at05} have already been 
performed to understand the effect of the gas sound speed on the gas flows inside 
the inner Lindblad resonance (ILR). The smoothed particle hydrodynamics (SPH)
simulations of \citet{eng97} and \citet{pat00} showed that the off-axis shocks 
form at low gas sound speed, while the on-axis shocks develop at high gas sound speed.
However, they were unable to  analyze the nuclear morphology in detail
due to the lack of resolution of their simulations.
On the other hand, the symmetric two-armed nuclear spirals generated by 
the gravitational torque of the bar were found in the high-resolution hydrodynamic 
simulations of \citet{eng00}, \citet{mac02}, and \citet{at05}. 
It was shown by \citet{eng00} with grid-based hydrodynamic simulations 
that the symmetric two-armed nuclear spiral can be explained 
in terms the gas density waves, since the spiral arms open out 
as the gas sound speed increases. A similar effect of the gas sound speed 
on the openness of symmetric two-armed nuclear spirals was also 
found in the SPH simulations of \citet{at05}. However, their simulations 
suggested that the symmetric two-armed nuclear spirals are formed 
by hydrodynamic spiral shocks driven by the bar. 
Furthermore, \citet{mac02} found that the gas sound speed plays 
a crucial role in the mechanism responsible for the formation of 
the symmetric two-armed nuclear spirals in barred galaxies. 
Their grid-based hydrodynamic simulations showed that 
the symmetric two-armed nuclear spirals are formed by hydrodynamic shocks 
driven by the bar when the gas sound speed is high ($c_{s}=20$ km s$^{-1}$), 
while for gas at a low sound speed ($c_{s}=5$ km s$^{-1}$), they are maintained
by gas density waves. It was noticed in the grid-based hydrodynamic simulations of
\citet{mac04} that like the gas sound speed, the bar strength can also affect 
the mechanism responsible for the formation of the symmetric two-armed nuclear 
spirals in barred galaxies.

Since the central mass concentration determines the shape 
of the gravitational potential in the nuclear regions of galaxies, 
it can also affect the morphology of symmetric two-armed nuclear spirals. 
This can be understood from the simulations of \citet{eng00} 
and \citet{ann04}, where they have mentioned that the openness of 
symmetric two-armed nuclear spiral arms depends on the central mass concentration. 
Our previous simulations \citep{at05} showed that the presence of 
a central supermassive black hole (SMBH), which can greatly affect
the central mass concentration, plays a significant role in the sense that
the symmetric two-armed nuclear spirals form when the mass of an SMBH 
is massive enough to remove the inner inner Lindblad resonance (IILR). 
In addition to this, \citet{mac04} noticed that a central massive black hole (MBH) 
of sufficient mass can allow the symmetric two-armed nuclear spirals 
to extend all the way to its immediate vicinity.

However, it is important to note here that the effect of the central 
mass concentration on the mechanism responsible for the formation of 
the symmetric two-armed nuclear spiral was not known from previous studies, 
whereas that of the gas sound speed was understood from the simulations 
of \citet{mac02}. Furthermore, it is worth mentioning that if we consider
the model galaxy whose bulge central density varies as $r^{-\gamma}$,
its mass would be more centrally concentrated as the cusp parameter $\gamma$ 
increases. Thus, the purpose of the present paper is to examine the effect of
the central mass concentration by varying the cusp parameter $\gamma$
on the mechanism responsible for the formation of the symmetric two-armed 
nuclear spirals in barred galaxies. For this purpose, we employ 
the SPH technique to solve the hydrodynamic equations for the gas flow 
in the model galaxies whose bulge central densities vary 
as $r^{-\gamma}$. We have also included the self-gravity of the gas 
in our calculations, since this can play a significant role 
in high-density regions such as shocks.

The remainder of this paper is organized as follows. In \S \ 2 we describe 
the model galaxy, model parameters, the gaseous disk model, and the numerical method. 
In \S \ 3 we present the results of our simulations, and the discussion 
of our results is given in \S \ 4. The conclusions are presented in \S \ 5.

\section {MODELS AND NUMERICAL METHODS}

\subsection{The Model Galaxy}

The model galaxy is assumed to be composed of four components: 
halo, disk, bulge, and bar. The properties (density, size and structure) of 
all these potential generating components are assumed to be invariant 
in time in the frame corotating with the bar.

The halo component, which gives rise to the flat rotation curve 
at outer radii, is assumed to have a logarithmic potential 
with finite core radius,
\begin{equation}
\Phi (r)_{halo} = {1\over 2} v_0^2 \ln(R_h^2+r^2) + const,
\end{equation}
where $R_h$ is the halo core radius and $v_0$ is the constant
rotation velocity at large $r$. 

The disk component is assumed to have the potential of 
 \citet{fre70} exponential disk,
\begin{eqnarray}
\nonumber \Phi (r)_{disk} &=& \pi G \Sigma _0 r \left[ I_0\left(r
\over 2R_d\right)
K_1\left(r \over 2R_d\right)\right. \\
&-&\left. I_1\left(r \over 2R_d\right)K_0\left(r
\over 2R_d\right)\right],
\end{eqnarray}
where $R_d$ is the disk scale length, $\Sigma _0$ is the central surface
density, and $I_0,~K_0, ~I_1$, and $K_1$
are the modified Bessel functions. The total disk mass is
simply $M_{disk}=2\pi \Sigma_0 R_d^2$. 

For the bulge component, we adopt the potential of the spherical 
$\gamma$-models of \citet{deh93},
\begin{equation}
\Phi (r)_{bulge} = \frac{GM_{bulge}}{(2-\gamma )R_{b}}\left[
\left(\frac{r}{r+R_{b}}\right)^{2-\gamma }-1\right], \textrm{for $\gamma \ne 2$}, 
\end{equation}
where $M_{bulge}$ is the total mass of the bulge, $R_b$ is the scale-length,
and $\gamma$ is  the cusp parameter that can have a value in between 
$0 \le \gamma < 3$. Its corresponding density varies as $r^{-\gamma}$ 
near the center and $r^{-4}$ at large radii, and hence possesses 
a central density core for $\gamma = 0$ and cusps for  $\gamma > 0$. 
This bulge with $\gamma = 1.5$ most closely resembles 
the de Vaucouleurs' $R^{1/4}$-profile \citep{deh93} 
that is known to better represent the bulges of real galaxies. Hereafter, 
the central mass concentration  will  be represented by $\gamma$, 
since the central mass concentration of the model galaxies increases 
with the cusp parameter $\gamma$. 

The bar is a triaxial component in three dimensions. However, our simulation
is restricted to the two-dimensional disk. Thus, we assume 
the following form of the gravitational potential proposed 
by \citet{lon92} for the bar component:
\begin{equation}
\Phi (r)_{bar} = {G M_{bar}\over 2a}
 \log\left({x-a+T_- \over x+a+T_+}\right),
\end{equation}
where $T_\pm=\sqrt{[(a\pm x)^2+y^2+b^2]}$, and $a$ and $b$ are
 the needle length and softening length, respectively.
 The parameter $a/b$ defines the elongation of the bar, which is
 proportional to the axis ratio $\tilde{a}/\tilde{b}$ of the flattened 
homogeneous ellipsoid bar of \citet{fre66} when $a/b \ge 2$ 
\citep[see Fig. 3 of][]{lon92}, where $\tilde{a}$ and $\tilde{b}$ 
are the major and minor axes of a flattened homogeneous ellipsoid bar.

\subsection {Model Parameters}

\clearpage
\begin{table*}
\begin{center}
{TABLE 1}\\
\vskip 0.2cm
{\small MODEL PARAMETERS}
\vskip 0.2cm
\begin{tabular}{lcc}\hline \hline
 &  $c_{s}$   &      \\ 
Model &    (km s$^{-1})$ & $\gamma$ \\ \hline
M1  &    10. &  0.0     \\ 
M2  &    10. &  0.5   \\
M3  &    10. &  1.0   \\
M4  &    10. &  1.5   \\
M5  &    15. &  0.0   \\
M6  &    15. &  0.5   \\
M7  &    15. &  1.0   \\
M8  &    20. &  0.0   \\
M9  &    20. &  0.5   \\
M10 &    20. &  1.0  \\
\hline
\end{tabular}
\end{center}
\end{table*}
\clearpage
We have considered ten models M1-M10, which are listed in Table $1$. 
All these models are defined according to different choices of the gas sound 
speed $c_{s}$ and the central mass concentration $\gamma$. In this paper, we are mainly 
interested to examine the effect of the central mass concentration $\gamma$
on the mechanism responsible for the formation of the symmetric two-armed 
nuclear spirals in barred galaxies when the gas sound speed is chosen to be 
$c_{s}=$ 10 km s$^{-1}$ (see  \S \ 3.2 and \S \ 3.3 ). Thus, apart from 
the models with $\gamma=0$, $0.5$, and $1$, an additional model 
with $\gamma=1.5$ is considered for the gas sound speed 
of $c_{s}=$ 10 km s$^{-1}$ (i.e., model M4). However, the models 
with $\gamma=0$, $0.5$, and $1$ are chosen for each of the remaining 
gas sound speeds of $c_{s}=15$ and $20$ km s$^{-1}$. 
Since the total visible mass of the model galaxies is assumed to be 
$M_{G} \sim 4 \times 10^{10}$, the masses of the gas, disk, bulge, 
and bar components in units of $M_{sc} = 2 \times 10^{11}  M_{\odot}$
are $M_{gas}=0.002$, $M_{disk}=0.104$, $M_{bulge}=0.054$, and $M_{bar}=0.04$, 
respectively. Thus, the mass distributions of the model galaxies 
are chosen to be similar to those of early-type barred galaxies ($\sim$SBa), 
since the disk-to-bulge mass ratio and the fractional mass of the bar 
are $M_{disk}/M_{bulge} = 2$ and $M_{bar}/M_{G} = 0.2$, respectively.
The scale lengths of the halo, disk, and  bulge are assumed  to be 
$R_{h}=15.0$ kpc, $R_{d}=3.0$ kpc, and $R_{b}=0.25$ kpc, respectively. 
Furthermore, the bar needle length $a$, the bar softening length $b$, 
the bar pattern speed $\Omega_{p}$, and the rotation velocity of the halo  
$v_{0}$ are fixed at $3.0$ kpc, $1.0$ kpc, $42.8$  km s$^{-1}$ kpc$^{-1}$, 
and  $186$  km s$^{-1}$, respectively. 

Figure $1$ shows the rotation curves and angular frequency curves of 
the models mentioned in Table $1$ that have the central mass concentrations of 
$\gamma=0$ and $1$. Since the models with constant values of $\gamma$ 
have the same mass distribution, we present the rotational curves 
and angular frequency curves of models M1 and M3 that assume $\gamma=0$ and $1$, 
respectively. The top left panel of Figure $1$ represents 
the rotational velocities of the halo, disk, bulge, and bar components 
of model M1 with $\gamma = 0$ as a function of radius, while the top right panel 
shows the rotational curves of model M3 with $\gamma = 1$. 
Since the bar is nonaxisymmetric component, its contribution 
to the rotational velocity is included here after averaging it axisymmetrically
\citep{has93}. As can be seen in the top panels of Figure $1$,  
the rotational velocities in the vicinity of the galactic center
are affected by the increase in central mass concentration
from $\gamma =0$ to $1$. However, the bottom panels of Figure $1$ 
represent the angular frequencies corresponding to the rotational velocity 
generated by the axisymmetric components as a function of radius. 
The adaption of $\Omega_{p}$ = $42.8$  km s$^{-1}$ kpc$^{-1}$ 
places the IILR at the negligible distance from the galactic center, 
and the outer inner Lindblad resonance (OILR) near $1.3$ kpc in model M1 
with $\gamma = 0$, whereas it places the single ILR close to $1.5$ kpc
in model M3 with $\gamma = 1$. Furthermore, the corotation radius ($R_{CR}$) 
is found to be around $1.2a$ in models M1 and M3 for the above choice of $\Omega_{p}$. 
Since the angular frequency curve reaches high values (larger than $\Omega_{p}$) 
at the nucleus for model M1 that considers the central mass concentration 
of $\gamma = 0$, it suggests that for the assumed values of $M_{bulge}=0.054$ 
and $R_{b}=0.25$ kpc, the mass of model M1 with $\gamma = 0$ is sufficiently 
centrally concentrated  \citep{mac03} to weaken the IILR by placing its location 
at the negligible distance from the galactic center. 
However, the IILR is completely removed and the angular frequency curve 
goes to infinity at the galactic center for model M3 with $\gamma = 1$, 
since the central mass concentration increases from $\gamma=0$ to $1$. 
Although it is not presented here, we would like to 
mention that except for the slight difference in the very vicinity 
of galactic center, the rotation curves of the models  with 
the remaining values of $\gamma$ = $0.5$ and $1.5$ appear to be similar 
to those of model M3 with $\gamma = 1$. In addition to this, 
the angular frequency curves of the models with $\gamma$ = $0.5$ 
and $1.5$ do not greatly differ from those of model M3 with $\gamma = 1$,
except for the minor shift in the location of the single ILR.

\subsection{The Gaseous Disk Model}

In the direction of modeling the response of the gaseous disk to the imposition 
of nonaxisymmetric bar potentials, many significant results have been produced 
so far when the gas is isothermal 
\citep[eg.,][]{eng97,fuk98,fuk00,fux99,fux01,lee99,pat00,eng00,ann01,mac02,
mac03,mac04,ann04,at05,per08,name08}. 
\citet{eng97}, \citet{eng00}, and \citet{mac03} pointed out 
that this approach of considering isothermal gas is consistent with the ISM model 
of \citet{cow80}, who argued that the cloud fluid can be treated like 
an isothermal gas if the clouds have an equilibrium mass spectrum. The equilibrium 
is assumed to be maintained by a steady supply of small clouds by supernovae. 
Thus, we assume the infinitesimally thin and isothermal uniform gaseous disks with 
the effective sound speeds of $c_{s} = 10$, $15$, and $20$ km s$^{-1}$. 
The mass of the gaseous disk was chosen to be $1\%$ of the total visible mass 
of the model galaxy, since we simulate the response of the central kiloparsec region of 
the gaseous disk in early-type barred galaxies. In order to represent 
the initial gaseous disk with constant density, $2 \times 10^4$ SPH particles 
are distributed uniformly over a grid inside a circular disk of 
$5$ kpc radius. The initial circular velocities are given to the SPH particles 
so that the centrifugal accelerations of the particles are balanced with 
their gravitational accelerations. In the context of this equilibrium 
gaseous disk model, the pressure gradient forces are zero. 
The initial resolution length, defined by equation (12), 
was found to be $\sim147$ pc. However, the resolution of the SPH simulations 
becomes far better than the initial values in the central region 
as the density in the nuclear region increases within 
a few bar rotation periods (see \S \ 2.4). We introduce the bar potential 
slowly within one-half bar rotation period to avoid a spurious response 
of the gaseous disk, where the bar rotation period 
$\tau_{bar}$ is $\sim 1.44\times 10^8$ yr. 
The previous simulations of the isothermal gas flow in barred galaxies
with broad spectrum of values for effective sound speed \citep{eng97,pat00,mac02} 
and the compilation of HI velocity dispersion by  \citet{kam93} 
suggest that our adopted values of $c_{s} = 10$, $15$, and $20$ km s$^{-1}$ 
correspond to the vertical cloud velocity dispersion in the Galactic bulge.

\subsection{Numerical Method}

We have used the PMSPH method to solve the hydrodynamical equations for 
the gas flow in barred galaxies. Our simulations are confined in 
two dimensions, since restricting the calculation to two dimensions increases 
greatly the resolution, which is important in the present problem of 
resolving the shocks \citep[cf.][]{eng97}. In this regard, \citet{eng97} 
mentioned that the two-dimensional description appears to be a reasonable 
approximation, since they have done some three-dimensional simulations, 
where the gas flows were indeed similar to the two-dimensional ones. 
We have used the PMSPH code of \citet{fux99,fux01}, which has recently
been used by \citet{per08} to model the gas flow in the barred galaxy
NGC 4123. R. Fux of Geneva Observatory Group has kindly provided 
the PMSPH code to us. Here we briefly describe the necessary specifications 
relevant to the particular aspects of our PMSPH code and refer the readers 
to \citet{fux99,fux01} for a detailed description of our code and 
to \citet{ben90}, \citet{mon92}, and \citet{ste96} for 
the general properties of solving Euler's equation of motion 
using the popular Lagrangian method. 

\subsubsection{Gas Hydrodynamics with SPH}

Since SPH  is fully Lagrangian, it does not have any grid to impose 
any artificial restrictions on the global geometry of the systems 
under study or any mesh-related limitations on the dynamic range 
in spatial resolution. Thus, a high numerical resolution 
is naturally achieved in high-density regions such as shocks 
by using SPH \citep[][]{fuk98}, which is necessary to solve 
our present problem of understanding the response
of gaseous disks in the nuclear regions of barred galaxies.

Using the SPH method, we solve the Euler equation of motion that can be written as
\begin{equation}
\frac {d {\bf v}} {dt} \equiv \frac {\partial {\bf v}} {\partial t} +
  ( {\bf v}  \cdot {\bf \nabla})
 {\bf v} = - \frac {{\bf \nabla} (P+\Pi)} {\rho} - {\bf \nabla} \Phi,
\end{equation}
where ${\bf v}({ \bf r})$ is the velocity field and $\rho ({\bf r})$,
$P({\bf r})$ and $\Pi({\bf r})$ are the density, pressure, and
artificial viscosity of the gas, respectively.  $-{\bf \nabla} \Phi$ is the
gravitational force, where $\Phi$ includes potentials of the gas,  
as well as  the stellar and dark components. 

In SPH, the density is evaluated directly at each particle position
${\bf r}_{i}$ in space by summing the contributions from  the density
profiles of neighboring particles and is represented  by
\begin{equation}
\rho_{i} \equiv \rho({\bf r}_{i}) = \sum_{j=1}^{N_{g}} m_{j} W({\bf r}_{i}
  - {\bf r}_{j}, h),
\end{equation}
where $N_{g}$ is the number of gas particles, $m_{j}$ is the mass of
individual particles, $W({\bf r}, h)$ is the kernel function, and $h$ is the
smoothing length that defines the local spatial resolution and is
proportional to the local mean interparticle spacing. In our code, 
the adopted SPH kernel is a spherically symmetric spline 
that vanishes outside $2h$ because of its finite spatial extension. 

Our code  assumes an isothermal equation of state, i.e.,
\begin{equation}
P_{i}= {c^{2}_{s}} \rho_{i},
\end{equation}
where $c_{s}$ is the effective sound speed, which can be interpreted 
globally as the velocity dispersion of the interstellar clouds. 
The internal energy of the gas is at any time and everywhere constant, 
and in particular, there is no energy equation involved in our code.
With the above equation (7), the pressure gradient term of \citet{ben90}
in the Euler's equation simply takes the form
\begin{equation}
\left(\frac {{\bf \nabla} P} {\rho} \right)_{i} \approx {c_{s}^2}
  \sum_{j=1}^{N_{g}} m_{j} \left(\frac {1} {\rho_{i}} + \frac {1} {\rho_{j}}
  \right) {\bf \nabla}_{i}  W({\bf r}_{i} - {\bf r}_{j}, h).
\end{equation}

The introduction of an artificial viscosity is required for an accurate 
treatment of the flow near shocks. In our code, the artificial viscosity
is calculated exactly as in \citet{ben90}. It consists of the bulk and 
the von Neumann-Richtmyer viscosities, with the standard parameters set 
to $\alpha=1.0$ and $\beta=2.5$, and takes into account \citet{bal95} 
correction to avoid energy dissipation in pure shearing flows.

In order to achieve high resolution, it is necessary to treat 
the smoothing length $h$ properly, since it determines the local spatial resolution.
To increase the resolution in high-density regions like shocks, 
our code uses a spatially variable smoothing length $h$ to assign 
an individual smoothing length $h_{i}$ to each particle in such a manner
that the  number of neighbor particles $N_{i}$ to each particle 
always remains as close as possible to a fixed number $N_{o}$.
This allows us to save significant computing time while 
keeping spatial resolutions high enough to resolve the shocks 
\citep[][]{fux99,fux01}. Two particles $i$ and $j$
are defined as mutual neighbors if $i \ne j$ 
and $\vert {\bf r}_{i} - {\bf r}_{j} \vert < 2 h_{ij}$,
where $h_{ij} =  (h_{i} + h_{j})/2$ is the symmetrized smoothing
length, which is supplied in equations $(6)$ and $(8)$, as well as  
in the formula for the artificial viscosity to ensure momentum conservation.

At each time step, the smoothing lengths are updated in our code
according to the general three-dimensional scaling law:
\begin{equation}
\frac {h_{i}} {h_{o}} = \left(\frac {N_{i} +1} {N_{o} +1} \frac
  {\rho_{o}} {\rho_{i}} \right)^{1/3},
\end{equation}
where ${h_{o}}$ and  ${\rho_{o}}$ are constants, $N_{i}$ is the number of
neighbors of particle $i$, and ${+1}$ is added 
to take into account particle $i$. Following \citet{ben90}, we take
the time derivative of equation $(9)$ and substitute the continuity equation 
to yield  
\begin{equation}
\dot{h_{i}} \equiv \frac {d h_{i}} {d t} = \frac {1} {3} h_{i}
  \left(\frac {1} {N_{i}+1} \frac {d N_{i}} {d t} + \lbrack
  {\bf \nabla} \cdot {\bf v} \rbrack_{i} \right).
\end{equation}
In equation $(10)$, \citet{ben90} did not include the term  ${d N_{i}}/{dt}$ to ensure 
a constant number of neighbors. However, this does not prevent 
a slow numerical departure of the $N_{i}$'s from $N_{o}$ with time.
To overcome this problem, our code simply damps such departures by
setting 
\begin{equation}
\frac {d N_{i}} {d t} = \frac {N_{i} - N_{o}} {\eta \triangle t}
\end{equation}
in equation $(10)$ and integrates the resulting equation along with the
equations of motion. The parameter $\eta$ controls the damping rate 
per time step $\triangle t$ and should be significantly
greater than $1$ to avoid abrupt discontinuities in nongravitational forces. 
In all our simulations, we have assumed $N_{o}=35$ and $\eta = 5$.

The initial smoothing lengths of the gas particles, confined near 
the plane $z=0$, are defined as 
\begin{equation}
h_{i}(t=t_{o}) = \sqrt \frac {N_{o} m_{i}} {4 \pi \Sigma(R_{g})},
\end{equation}
where $m_{i}$ and  $\Sigma(R_{g})$ are the mass of each gas particle
and the surface density of the initial gaseous disk of 
radius $R_{g} = 5$ kpc, respectively.

It is customary to impose quite a large lower limit on the minimum 
smoothing length $h_{min}$ to save computing time. However, because of 
the fast neighbor searching algorithm adopted in our code, we have used 
$h_{min}$ = 0.1 pc to resolve structures down to the parsec scale. 
Although the initial smoothing length (i.e., the initial resolution length) 
of $\sim 147$ pc is employed in our simulations (see \S \ 2.3), 
the smallest smoothing length actually achieved after a few bar rotation periods 
is $\sim 2$ pc, which is smaller than the $h_{min}=5$ pc used in 
the high-resolution SPH simulations of \citet{fuk00}. We find that
the  smoothing length of $\sim 2$ pc is small enough 
to resolve the nuclear spirals, as shown in \S \ 3. 
Furthermore, it is worth mentioning here that our code uses 
a synchronized version of the standard leap-frog time integrator 
\citep[e.g.,][]{hut95}, which is known to well conserve 
the total energy and the total angular momentum and also employs 
an adaptive time step to temporally resolve the high-density shocks 
in the gaseous disk \citep[see][]{fux99,fux01}.

\subsubsection{Gravitation with PM}

Our gaseous disk is not a self-gravitating one, since its mass is assumed
to be only a tiny fraction of the model galaxy ($\sim 1 \%$). This suggests
that Toomre's $Q$ value is much greater than unity (i.e., $Q>>1$), which represents 
a stable gaseous disk \citep[cf.][]{elm93,fuk98,fuk00,at05}. However, we have 
included the self-gravity of gas in our calculations because it can play 
a significant role in high-density regions such as shocks that develop  
in the later evolution of gaseous disk. To compute to the self-gravity of gas, 
our code uses the particle-mesh (PM) technique with polar-cylindrical grid geometry 
described in \citet{pfe93}. The advantage of considering the polar-cylindrical grid 
is that the radial and azimuthal resolutions increase toward the center, where a
variety of features are likely to develop because of the gas inflow driven by the bar. 
Using the convolution theorem, the potential is computed on the polar-cylindrical grid 
by the fast Fourier transform (FFT) technique in the $\phi$ and $z$ dimensions, 
where the cells are equally spaced. The radial spacing of the cells is logarithmic with 
a linear core to avoid an accumulation point at the center. 
Since the Keplerian kernel does not truly satisfy the conditions
of the convolution theorem, a variable homogeneous ellipsoidal
kernel is adopted in our code for the softening of the short range forces 
\citep[cf.][]{pfe93,fux99,fux01}. This kernel tends toward the Keplerian kernel
at large distance, and can also be useful in situations that
need anisotropic resolution \citep[][]{pfe93}. 
Since the polar-cylindrical grid is non-homogeneous, a variable softening length 
is necessary. Otherwise a softening length intermediate between the smallest and 
largest cell sizes both decreases resolution in some parts 
of the grid and enhances grid noise in other parts of the grid. 
Thus, the softening lengths represented by the semi-axes of 
our adopted variable homogeneous ellipsoidal kernel have been set 
as $1.1$ times the respective cell dimensions. This ensures 
the conservation of the total energy, as well as of the total angular momentum
while keeping high central resolution \citep{pfe93,fux99,fux01}. 
For a more detailed description of the PM technique used in our code, 
we simply refer the readers to \citet{pfe93} and \citet{fux99,fux01}.

\section {RESULTS}

\subsection {Nuclear Feature of the Gaseous Disk}

The nuclear features of the gaseous disks in models M1-M10 at the evolution time
of $10 \tau_{bar}$ are shown in Figure $2$. Here our results are shown on 
the frame corotating with the bar, which always lies horizontally. 
As can be seen in Figure $2$, models M1-M10 clearly show the development of 
symmetric two-armed nuclear spirals at the evolution time of $10 \tau_{bar}$. 
For models M1-M4 in the first column that consider the gas sound speed
of $c_{s}=10$ km s$^{-1}$, the innermost parts of the symmetric two-armed nuclear spiral 
of model M1 with $\gamma=0$ extend up to winding angle of $\sim \frac {9} {4} \pi$,
whereas they can wind up to $\sim 2 \pi$ for model M4 with  $\gamma=1.5$.
It is also apparent that the nuclear disks are completely filled 
with the gas particles for models M8-M10 in the last column 
that assume the gas sound speed of $c_{s}=20$ km s$^{-1}$. 
In this case, the innermost parts of the symmetric two-armed nuclear spiral 
of model M8 with $\gamma=0$ extend all the way to the center 
with winding angle of $\sim 3 \pi$, whereas they appear to be 
tightly wound in the sense that their arms cannot be resolved 
after $\sim \frac {5} {2} \pi$ for model M10 with  $\gamma=1$.
These characteristics of the nuclear features show that 
the symmetric two-armed nuclear spiral arms open out as the central 
mass concentration $\gamma$ decreases while keeping the gas sound speed $c_{s}$ constant. 
By comparing the winding angles corresponding to the symmetric two-armed  
nuclear spirals of the models with constant values of $\gamma$ 
in Figure $2$ such as those mentioned above for models M1 and M8 that assume 
$\gamma=0$ but different gas sound speeds of $c_{s}=10$ and $20$ km s$^{-1}$, 
respectively, it is clear that the nuclear spiral arms also open out 
as the gas sound speed  $c_{s}$ increases while keeping the central 
mass concentration $\gamma$ constant. In addition to the above  
nuclear features of the gaseous disks in models M1-M10, Figure $2$ also shows 
that the diameter of the nuclear disk depends on the central mass concentration 
$\gamma$ and the gas sound speed $c_{s}$. The models with high central mass 
concentrations allow a larger nuclear disk than those with low central 
mass concentrations. Furthermore, a larger nuclear disk is also 
found in the models with low gas sound speeds. 
As a result, a largest nuclear disk among our selected models 
is noticed in model M4, which assumes $\gamma=1.5$ and $c_{s}=10$ km s$^{-1}$.

\subsection {Nuclear Distribution of the Shocked Gaseous Particles}

As shown in the previous section, all the models in the present study  show 
the formation of the symmetric two-armed nuclear spirals at the evolution 
time of $10 \tau_{bar}$. The mechanisms such as the hydrodynamic shocks 
\citep{mac02, at05} and the gas density waves \citep{eng00} were proposed 
for the formation of such nuclear spirals in barred galaxies. 
Since a significant role is played by the gas sound speed $c_{s}$ 
to decide whether the symmetric two-armed nuclear spirals 
are formed by hydrodynamic shocks or by gas density waves \citep[cf.][]{mac02}, 
we are interested here to examine any such role of the central mass 
concentration $\gamma$. For this study, we mainly consider models M1-M4, 
which have $c_{s}=10$ km s$^{-1}$ but different central mass concentrations 
of $\gamma$ = 0, 0.5, 1, and 1.5, respectively. In order to see whether 
the symmetric two-armed nuclear spirals shown in Figure $2$ 
for models M1-M4 are made of shocked gas particles, we examine 
the distributions of gas particles that are shocked by supersonic flow, 
i.e., $-h {\bf \nabla} \cdot {\bf \upsilon} > c_{s}$ \citep{eng97,at05}, 
in the central kiloparsec of models M1-M4 at the evolution 
time of $10 \tau_{bar}$ in Figure $3$. As can be seen in Figure $3$,
the distributions of shocked gas particles in the central kiloparsec 
of models M1 and M2 reveal well-defined spiral patterns whose innermost 
parts extend up to winding angle of $\sim 2 \pi$. These nuclear spiral shocks look 
very similar to the symmetric two-armed nuclear spiral patterns 
developed in Figure $2$ for the distributions of all the gas particles 
in the central kiloparsec of models M1 and M2. 
On the other hand, it is also apparent from Figure $3$ that 
the distributions of shocked gas particles in the central kiloparsec 
of models M3 and M4 cannot move inward in a curling manner 
after winding angle of $\pi$. However, the distributions of all the gas particles in 
the central kiloparsec of models M3 and M4 show the development of
symmetric two-armed nuclear spirals whose innermost parts extend up 
to winding angle of $\sim 2 \pi$, as shown in Figure $2$.
Thus, our simulations suggest that the symmetric two-armed nuclear spirals 
of models M1 and M2 that respectively assume $\gamma=0$ and $0.5$ are formed 
by hydrodynamic spiral shocks driven by the gravitational 
torque of the bar, whereas they cannot be supported by the shocked gas particles
for models M3 and M4 that respectively assume $\gamma=1$ and $1.5$. 
It is worthwhile to mention here that the models with each of the remaining values of
the gas sound speed (i.e., $c_{s}=15$ and $20$ km s$^{-1}$)
show a similar effect of the central mass concentration  $\gamma$
on the formation of the nuclear spirals in barred galaxies 
to that shown in Figure $3$ for the models with $c_{s}=10$ km s$^{-1}$.

\subsection {Effect of Central Mass Concentration on Gas Inflow}

Symmetric two-armed nuclear spirals are suggested to be formed by 
the gas inflow across the ILR due to the loss of angular momentum at 
the shocks driven  by the gravitational torque of the bar 
\citep{mac03,mac04,at05}. Thus, it is quite plausible that 
the central mass concentration $\gamma$ can affect the gas inflow 
into central hundred parsec of models M1-M4 with  $c_{s}=10$ km s$^{-1}$,
since Figure $3$ shows that the central mass concentrations of $\gamma=0$ 
and $0.5$ make physical conditions in which it is plausible 
for the symmetric two-armed nuclear spirals to take the form of hydrodynamic shocks. 
This inspires us to examine the effect of the central mass concentration $\gamma$ 
on the gas inflow into the central hundred parsec of models M1-M4 in detail. 
In order to do this, the time evolution of the fraction of gas particles 
accumulating inside the radial zone of $0.3$ kpc is shown in Figure $4$ 
for models M1-M4. The gas accumulated inside the radial zone of $0.3$ 
represents the gas inflow deep inside the ILR, since the OILR in model M1 
and the ILRs in models M2-M4 are located around at $1.3$ kpc and $1.5$ kpc, respectively. 
As can be seen in Figure $4$, models M1-M4 show appreciable gas inflow into the 
central hundred parsec throughout the evolution. Although there are 
little differences in the gas inflow among the models in the early phase 
of evolution, model M4 with $\gamma=1.5$ shows slightly larger gas inflow
up to $\sim 6 \tau_{bar}$. However, at the later evolution times 
(i.e., after $\sim 6 \tau_{bar}$), the remarkable differences in 
the gas inflow are found in the sense that the gas inflow increases 
as the central mass concentration decreases from $\gamma = 1.5$ to $0$. 
As a result, the highest gas inflow rate is found for model M1 
with $\gamma=0$ whose mass is less centrally concentrated than models M2-M4
that assume $\gamma=0.5$, $1$ and $1.5$, respectively. 
This correlation between the amount of gas inflow close to the center 
and the central mass concentration still exists when the models 
with each of the remaining values of the gas sound speed 
(i.e., $c_{s}=15$ and $20$ km s$^{-1}$) are considered. 
However, similar to \citet{pat00} and \citet{at05}, 
the gas inflow rate increases when the gas sound speed 
increases from  $c_{s}=10$ to $20$  km s$^{-1}$.

\section {DISCUSSION}

\subsection {Dependence of Gas Flow Morphology on 
Central Mass Concentration and Gas Sound Speed}

Using the SPH method, \citet{eng97} have found that 
when the gas in barred galaxies is modelled as an isothermal 
fluid, the gas flow morphology depends on the shape of 
the underlying potential and the sound speed. They 
noticed that the off-axis principal shocks with a nuclear ring 
developed by the gas flow at low sound speed turn into
the on-axis ones with a straight inflow to the center
when the sound speed is high. However, the resolutions of 
their simulations were not high enough to analyze the
nuclear morphology in detail. Our present SPH simulations 
confirm their findings that not only the underlying 
potential, which can be greatly affected by the central mass
concentration, but also the gas sound speed influence 
the morphology of the gas flow. 

Because of the better resolution of our SPH simulations, 
we analyze the gas flows in the central kiloparsecs of
barred galaxies, where the morphology of nuclear spirals 
is found to be dependent on the central mass concentration $\gamma$ 
and the gas sound speed $c_{s}$ (see \S \ $3.1$).
The openness of the nuclear spiral arms increases 
as the central mass concentration decreases. 
Increasing the gas sound speed has a comparable effect on 
the openness of nuclear spirals to that of decreasing 
the central mass concentration. This implies that 
the nuclear spiral arms also open out when the gas sound speed 
increases. As a result, the less centrally concentrated model with 
high gas sound speed (i.e., model M8 with $\gamma=0$ 
and $c_{s}=20$ km s$^{-1}$) shows more loosely wound nuclear spiral 
whose openness appears to be larger than those of the other selected models.
This can be justified by the fact that the innermost parts of
the nuclear spiral formed in model M8 extend all the way to 
the galactic center with winding angle of $\sim 3 \pi$ 
(see \S \ $3.1$ and Fig. $2$). A similar effect of 
the central mass concentration and the gas sound speed 
on the morphology of nuclear spirals was found by \citet{eng00}, 
although they used the two-dimensional grid-based 
hydrodynamics code ZEUS-2D. In addition to this, \citet{mac02},
who use the grid-based hydrodynamics code, have also found 
that the nuclear spiral opens out in the sense that its innermost parts  
reach very close to the galactic center when the gas sound 
increases from $c_{s} = 5$ to $20$ km s$^{-1}$.

\citet{at05}, who use the SPH code, have found that the models with
lower gas sound speeds  allow a larger diameter of the nuclear disk 
than those with high gas sound speeds. Here we have also found a similar effect 
of the gas sound speed on the diameter of the nuclear disk 
(see  \S \ $3.1$ and Fig. $2$). Beside this, we have noticed that 
the diameter of the nuclear disk increases as the central mass 
concentration $\gamma$ increases. A similar effect of the central mass 
concentration and the gas sound speed on the diameter of the nuclear disk 
has been found by \citet{eng00}. Thus, it is apparent that increasing 
the central mass concentration has a similar effect on the diameter 
of the nuclear disk to that of decreasing the gas sound speed.

Furthermore, it was also found by \citet{at05} that the nuclear regions 
of the modeled gaseous disks evolve to trailing nuclear spirals 
whose morphologies depend on the gas sound speed in the presence of 
a central SMBH that is massive enough to remove the IILR, 
whereas the lower gas sound speed model ($c_{s} = 10$ km s$^{-1}$) 
that has the existence of the IILR due to
the absence of a central SMBH evolves to a leading nuclear spiral. 
In general, the leading nuclear spirals are generated at the IILR 
because of the positive torque of the bar \citep{com02}, 
while the trailing nuclear spirals at the OILR 
(or at the ILR when there is no IILR) are made by orbital switching 
from the $x_{1}$-orbits that align with the bar axis to the $x_{2}$-orbits
perpendicular to the bar \citep{ath92,wad94,knp95,eng00}. 
We have not seen here any such existence of the leading nuclear spirals, 
since the masses of models with $\gamma=0$ are sufficiently centrally 
concentrated to weaken the IILR by placing its location at the negligible 
distance from the galactic center (see Fig. $1$). This makes physical conditions 
plausible for the development of trailing nuclear spirals in the models 
with $\gamma=0$ whose innermost parts can reach the galactic center 
when the gas sound speed is high (see Fig. $2$). A similar dependence 
of the morphology of trailing nuclear spirals on the gas sound speed 
was found for the models with $\gamma \ge 0.5$ that have only a single ILR 
due to the increase of central mass concentration from $\gamma=0$. 

The relative bar strength is important in galaxy morphological studied 
because phenomena such as the gas inflow can be affected in various 
ways to the effectiveness with which the bar potential influences 
the motion of gas in a galactic disk \citep[eg.,][]{sell93,buta96,knp99}.
As mentioned in \citet{blk01}, the  relative bar strength defined by 
the maximum of the tangential-to-radial force ratio $(F_{Tan}/F_{Rad})_{max}$
increases when the bulge becomes weak.
We have found here that the gas inflow into the central hundred parsec 
increases as the central mass concentration decreases from $\gamma=1.5$ 
to $0$ (see \S \ $3.3$ and Fig. $4$). This correlation between the amount 
of gas inflow close to the center and the central mass concentration 
can be understood if we consider the dependence of the relative bar strength 
$(F_{Tan}/F_{Rad})_{max}$ on the strength of the bulge.
Since the relative bar strength $(F_{Tan}/F_{Rad})_{max}$ increases as the bulge 
becomes weak with the decrease in central mass concentration $\gamma$, 
the gravitational torques of the bars would be stronger in the models 
with $\gamma=0$ than in the models with $\gamma=1.5$. 
This provides favorable conditions for the gas particles in the models 
with $\gamma=0$ to lose their angular momentum much faster than those 
in the models with $\gamma=1.5$. Thus, the highest gas inflow rate is 
found for model M1 with $\gamma=0$ (see \S \ $3.3$ and Fig. $4$), 
since the gas particles move inward because of the loss of angular momentum. 
Apart from the relative bar strength, the gas inflow close to the center
can also be understood by the ILR that exists only if the central mass concentration
is large enough \citep{eng97,pat00,eng00}. The ILR is defined as the outer limit 
to which the $x_{2}$-orbits can extend \citep{mac04}. If the ILR is absent, 
the $x_{2}$-orbits do not exist, and strong gas inflow to the center 
can take place \citep{ath92,pin95}. As can be seen in Figure $2$, 
decreasing the central mass concentration from $\gamma=1.5$ to $0$ corresponds 
to the decrease in the strength of the ILR in the sense that the diameter
of the nuclear disk, which represents the region occupied by the $x_{2}$-orbits, 
decreases \citep[cf.][]{eng00}. This causes the highest gas inflow rate 
in the less centrally concentrated model (eg., model M1 with $\gamma=0$), 
as suggested by \citet{eng00}.

Using the SPH method, \citet{pat00} and \citet{at05} have found
that the gas inflow inside the radial zone near the central regions
increases with the gas sound speed. In addition to this, some previous simulations,
which either use the SPH code \citep{eng97} or the grid-based hydrodynamics
code \citep{eng00,mac02,mac04}, also indicate that the gas inflow close 
to the center increases with the gas sound speed. Although we have only presented here
the effect of the central mass concentration on the gas inflow, 
a closer look at Figure $2$ suggests a similar correlation 
between the amount of gas inflow  close to the center and 
the gas sound speed to that found in previous simulations. 
This can be understood if we consider the findings of \citet{pat00} 
and \citet{eng00} that while the gas particles
in the low sound speed mediums easily follow the $x_{2}$-orbits, 
the gas particles in the high sound speed mediums face more difficulty doing so.
Because of this reason, Figure $2$ shows that the  diameter of the nuclear disk, which 
represents the region occupied by the $x_{2}$-orbits, decreases when the gas 
sound speed increases from $c_{s} = 10$ to $20$ km s$^{-1}$ \citep[cf.][]{eng00}, 
resulting in a large amount of gas inflow deep inside the ILR for the models
with high gas sound speed ($c_{s} = 20$ km s$^{-1}$). Thus, the above described 
correlations make physical conditions in which it is plausible for 
the innermost parts of nuclear spiral to reach very close to the center 
when the less centrally concentrated model with high gas sound speed is 
considered (see Fig. $2$).

\subsection {Mechanism Responsible for the Formation of Nuclear Spirals}

The gas density waves \citep{eng00,mac02} and the hydrodynamic shocks 
\citep{mac02,at05} are two promising mechanisms proposed for the formation 
of the symmetric two-armed nuclear spirals in barred galaxies. 
As mentioned in \citet{eng00} and \citet{mac02}, the linear theory 
of gas density wave postulates the following: (1) the spiral pattern opens 
out as the gas sound speed increases; (2) the spiral pattern does not extend 
beyond the IILR when two ILRs are present; (3) when one ILR is present, 
the spiral pattern can exist between the center and the ILR. 
Since the gas density waves cannot propagate in regions where shocks 
are strong enough to drive a nonlinear response of gas flows, it can be applied to
low-amplitude spirals where no shocks exist and the gas response 
can be considered a linear wave  \citep{eng00,mac02}. 
However, it is worthwhile to mention that the gas density wave 
itself is triggered by the spiral shocks driven by the gravitational 
torque of the bar \citep{eng00}. 

The grid-based hydrodynamic simulations of \citet{mac02} show 
that the gas sound speed plays a significant role to decide whether the nuclear
spirals are formed by gas density waves or by hydrodynamic shocks.
In their low gas sound speed run ($c_{s}=5$ km s$^{-1}$), the nuclear spiral
is weak and rather tightly wound that creates at most a factor of $2$ 
density enhancement over the interarm region. Moreover, it terminates 
just outside the IILR, and its shock strength, as measured by div$^{2} {\bf v}$
for div ${\bf v} < 0$, is $10$ times smaller than that of the principal shock. 
This suggests that the nuclear spiral in their low gas sound speed 
run ($c_{s}=5$ km s$^{-1}$) can be well understood in terms of 
the gas density waves. 
On the other hand, the strong shock nature of nuclear spiral 
is found by \citet{mac02} in their high gas sound speed
run ($c_{s}=20$ km s$^{-1}$). They have found that the strong loosely wound 
nuclear spiral is a direct continuation of the principal shock 
along the leading edge of the bar, which generally agrees with the 
linear theory of gas density wave that postulates a open nuclear spiral for
a high gas sound speed \citep[cf.][]{eng00}. Nevertheless, the nuclear spiral 
reaches close to the nucleus well beyond the IILR where the gas density waves 
cannot penetrate. In addition to this, the  shock strength (i.e., div$^{2} {\bf v}$
for div ${\bf v} < 0$) and the arm/interarm density contrast of the nuclear spiral 
are comparable to those of the principal shocks. This indicates that the nuclear spiral 
in their high gas sound speed run ($c_{s}=20$ km s$^{-1}$) is definitely well beyond
the linear regime explored in the density wave theory where sound waves
cannot propagate in regions for which shocks are strong. Our previous SPH simulations
in the high sound speed mediums \citep{at05} also show the strong shock nature 
of nuclear spirals. In these simulations, we have found that 
the symmetric two-armed nuclear spirals are formed by hydrodynamic 
shocks in the presence of a central SMBH that can remove IILR when
the gas sound speed is high enough to drive a large amount of 
gas inflow deep inside the ILR.
 
Like the role of gas sound speed as described above \citep[cf.][]{mac02,at05}, 
we have found here that the central mass concentration also play a significant role
in the mechanism responsible for the formation  of the symmetric two-armed 
nuclear spirals in barred galaxies.
Our present SPH simulations show that the symmetric two-armed nuclear spiral arms 
open out as the gas sound speed increases (see \S \ $3.1$ and \S \ $4.1$), 
which generally agrees with the predictions of the linear theory of gas density wave.
Nevertheless, the symmetric two-armed nuclear spirals of the models with
$\gamma=0$ and $0.5$ cannot be explained by gas density waves, since they are
supported by the gas particles that are shocked by supersonic flow 
(see \S \ $3.2$ and Fig. $3$). On the other hand, since the symmetric two-armed 
nuclear spirals of the models with $\gamma=1$ and $1.5$ cannot be explained
by the shocked gas particles (see \S \ $3.2$ and Fig. $3$) and their arms 
open out when the gas sound speed increases, they can be well
understood in terms of the gas density waves. This suggests that
the symmetric two-armed nuclear spirals formed by hydrodynamic shocks 
turn into those explained by gas density waves when the central 
mass concentration increases from $\gamma = 0$ to $1.5$. 
Since the relative strength of bars in the models with $\gamma=0$ 
is stronger than that in the models with $\gamma=1.5$ (see \S \ $4.1$), 
our results appear to confirm the findings of the grid-based hydrodynamic
simulations of \citet{mac04} that the nuclear spirals developed 
in the models with strong bars have the nature of shocks in gas,
whereas those formed in the models with weak ovals can be explained 
in terms of the gas density waves. In addition to this, \citet{mac04} 
seems to support our findings that the nuclear spirals formed by hydrodynamic 
shocks trigger a large amount of gas inflow than those explained in terms 
of the gas density waves.

Although our present SPH simulations and the grid-based 
hydrodynamic simulations of \citet{eng00} show a similar
effect of the central mass concentration and the gas sound speed
on the morphology of nuclear spirals (see \S \ $4.1$), the shock nature of 
nuclear spiral was not found by \citet{eng00}. They have found 
the low-amplitude nuclear spirals, which can be
explained by gas density waves because the spiral arms open out
as the gas sound speed increases. On the other hand, we have 
found here that the nuclear spirals can be formed either by 
hydrodynamic shocks or by gas density waves, depending on 
the value of central mass concentration $\gamma$.
It appears that the different assumptions of the mass models 
along with the adoption of the different hydrodynamics codes 
might play some role in creating this discrepancy between 
our present simulations and those of \citet{eng00}.

\subsection {Effect of Number of Particles}

After performing the SPH simulations with $2 \times 10^{4}$ 
and $1 \times 10^{5}$ SPH particles for the response of 
the gaseous disk to the imposition of nonaxisymmetric bar potentials, 
\citet{pat00} claimed that the morphology of gas inflow does not 
depend on the number of SPH particles for $N_{g}$ larger than $2 \times 10^{4}$. 
This encourages us to check whether the number of SPH particles causes any major 
changes in our results discussed in the previous sections. Thus, we have repeated 
our simulations for models M1-M10 with $1 \times 10^{6}$ SPH particles 
that give the initial resolution of $\sim 20$ pc. Since we are mainly interested 
to examine whether the number of SPH particles affects our results discussed 
in \S \S \ 3.2 and 4.2 regarding the role of the central mass concentration 
$\gamma$ in the mechanism responsible for the formation of the nuclear spirals
in barred galaxies, we have presented here the nuclear distributions of supersonically 
shocked gas particles (i.e., $-h {\bf \nabla} \cdot {\bf \upsilon} > c_{s}$) 
for models M1-M4 with $1 \times 10^{6}$ SPH particles at the evolution time 
of $10 \tau_{bar}$ in Figure $5$. Similar to Figure $3$ that depicts the nuclear 
distributions of supersonically shocked gas particles for models M1-M4 
with $2 \times 10^{4}$ SPH particles at the evolution time of $10 \tau_{bar}$, 
Figure $5$ also shows that the nuclear spirals formed in models M1 with $\gamma=0$ 
and M2 with $\gamma=0.5$ have the nature of hydrodynamic shocks in gas, 
whereas those developed in models M3 with $\gamma=1$ and 
M4 with $\gamma=1.5$ are not supported by the shocked gas particles.
Although it is not presented in this paper, the results of our simulations,
which employ $1 \times 10^{6}$ SPH particles, are similar to those obtained 
in the previous sections using $2 \times 10^{4}$ SPH particles 
for the effect of the central mass concentration $\gamma$ and 
the gas sound speed $c_{s}$ on the morphology of nuclear spirals 
(see \S \ \S \ 3.1 and 4.1), as well as on the gas inflow close 
to the center (see \S \ \S \ 3.3 and 4.1). This implies that similar to 
our simulations with $2 \times 10^{4}$ SPH particles (see \S \ \S \ 3.1 and 4.1), 
the nuclear spiral arms formed in all the models with $1 \times 10^{6}$ 
SPH particles also open out as the gas sound speed $c_{s}$ increases. 
Because of this fact, our simulations with $1 \times 10^{6}$ SPH particles 
suggest that the nuclear spirals of models M3 with $\gamma=1$ and 
M4 with $\gamma=1.5$ can be explained in terms of the gas density waves 
\citep[cf.][]{eng00}, which is consistent with the formation mechanism 
proposed in \S \ 4.2 for the nuclear spirals of these models 
with $2 \times 10^{4}$ SPH particles. As a result, we do not see any changes in 
our findings regarding the effect of the central mass concentration $\gamma$ 
on the mechanism responsible for the formation of the nuclear spirals 
in barred galaxies when the number of SPH particles increases from 
$2 \times 10^{4}$ to $1 \times 10^{6}$. From this, we can
conclude that similar to \citet{pat00}, our results do not depend 
on the number of SPH particles for $N_{g}$ larger than  $2 \times 10^{4}$, 
and thus that this number of SPH particles is sufficient to examine 
the shock nature of nuclear spirals in barred galaxies.

\section{CONCLUSIONS}

We have used the SPH method to find the response of the nuclear region 
of a gaseous disk to the imposition of nonaxisymmetric bar potentials in order to
understand the effect of the central mass concentration $\gamma$ 
on the mechanism responsible for the formation of the symmetric two-armed 
nuclear spirals in barred galaxies. We have found that the central mass 
concentration $\gamma$ plays a significant role in shaping the gravitational 
potential in the central kiloparsec to weaken or to remove the IILR 
that prevents the gas inflow close to the nucleus. It was also noticed
that the gas flow driven by the gravitational torque of the bar leads to 
the formation of  symmetric two-armed nuclear spiral whose morphology depends  
on the central mass concentration $\gamma$ and the gas sound speed $c_{s}$ 
in the sense that the nuclear spiral arms open out when either 
the central mass concentration decreases or the gas sound speed increases.
Thus, the less centrally concentrated model with high gas sound speed  
shows more loosely wound nuclear spiral whose innermost parts reach
the galactic center in a curling manner. A largest nuclear disk 
is found for the more centrally concentrated model with low gas sound speed.

Furthermore, the central mass concentration $\gamma$ is found to play 
a crucial role to decide whether the symmetric two-armed nuclear spirals 
are formed by hydrodynamic shocks or by gas density waves.
The symmetric two-armed nuclear spirals are formed by 
hydrodynamic spiral shocks caused by the gravitational torque of the bar 
for the models with $\gamma = 0$ and $0.5$, whereas they are explained
in terms of the gas density waves when the models with $\gamma=1$ and $1.5$
are considered. This suggests that as the central mass concentration 
increases from $\gamma = 0$ to $1.5$, the mechanism responsible 
for the formation of the symmetric two-armed nuclear spirals 
in barred galaxies shifts from the hydrodynamic shocks to 
the gas density waves. The symmetric two-armed nuclear spirals 
formed by hydrodynamic shocks in the models with $\gamma=0$ 
drive a large amount of gas inflow than those explained by 
gas density waves in the models with $\gamma=1.5$.
Finally, we conclude that $2 \times 10^{4}$ SPH particles
give the same results as $1 \times 10^{6}$ SPH particles,
and thus are adequate to examine the shock nature of
nuclear spirals in barred galaxies.

\acknowledgments 
We thank the anonymous referee for useful remarks and 
suggestions that improve the present paper enormously.
We wish to express our sincere thanks to Roger Fux, who provided the PMSPH code. 
PT would like to express his sincere thanks to National Science 
Council (NSC), Taiwan, for granting postdoctoral fellowship through
grant: NSC 97-2811-M-007-017. PT and HBA are also thankful to  ARCSEC 
(Astrophysical Research Center for the Structure and Evolution of the Cosmos, 
Seoul, Korea) for providing support. HBA thanks Hyesung Kang for valuable 
discussion and comments on the numerical simulations. Most of the computations 
were conducted by  PC Cluster located at Department of Physics 
and Institute of Astronomy, National Tsing-Hua University, Hsinchu, Taiwan.
We are also grateful to the National Center for High-performance Computing (NCHC) 
for computer time and facilities.

\clearpage

\begin{figure}
\plotone{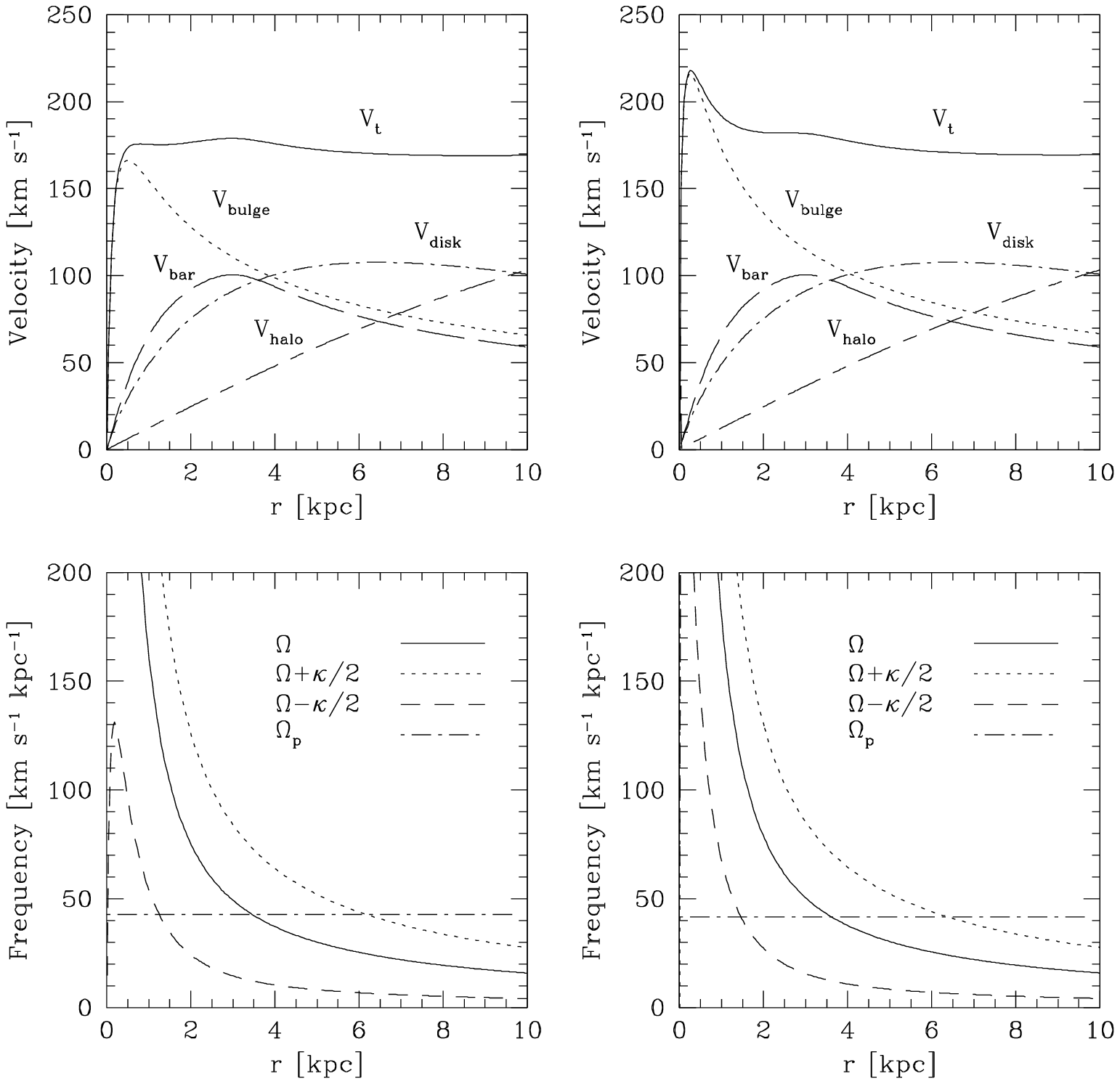}
\caption{Rotation curves and angular frequencies of model
galaxies. Because of the non-axisymmetric nature of the bar, its
 contribution is included here after averaging it axisymmetrically.
Left: Model M1 with $\gamma=0$; right: model M3 with
$\gamma=1$. The horizontal dot-dashed lines in the bottom panels 
represent the  bar pattern speed of $\Omega_{p}$ = 42.8 km s$^{-1}$ kpc$^{-1}$. }
\end{figure}
\clearpage
\begin{figure}
\plotone{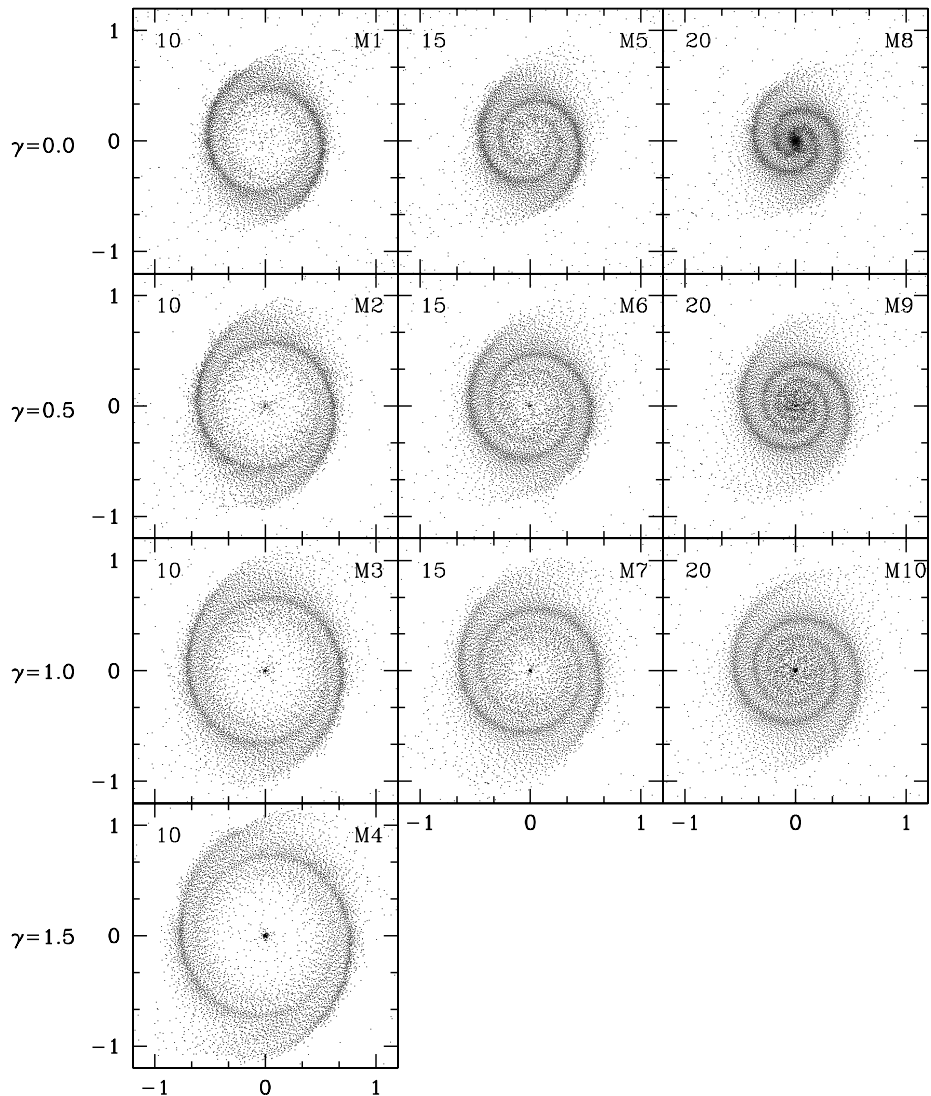}
\caption{Nuclear features of the gaseous disks in models M1-M10 
at the evolution time of $10 \tau_{bar}$. The bar rotation period 
$\tau_{bar}$ is $\sim 1.44\times 10^8$ yr. The models are indicated  in
the top right corner and the gas sound speed in units of kilometers per second
is given in the top left corner of each panel.
On the left side of each row, its corresponding value of the central mass 
concentration $\gamma$ is given. The bar lies horizontally, 
and the box size is $2.4$ kpc in one dimension.}
\end{figure}
\clearpage
\begin{figure}
\plotone{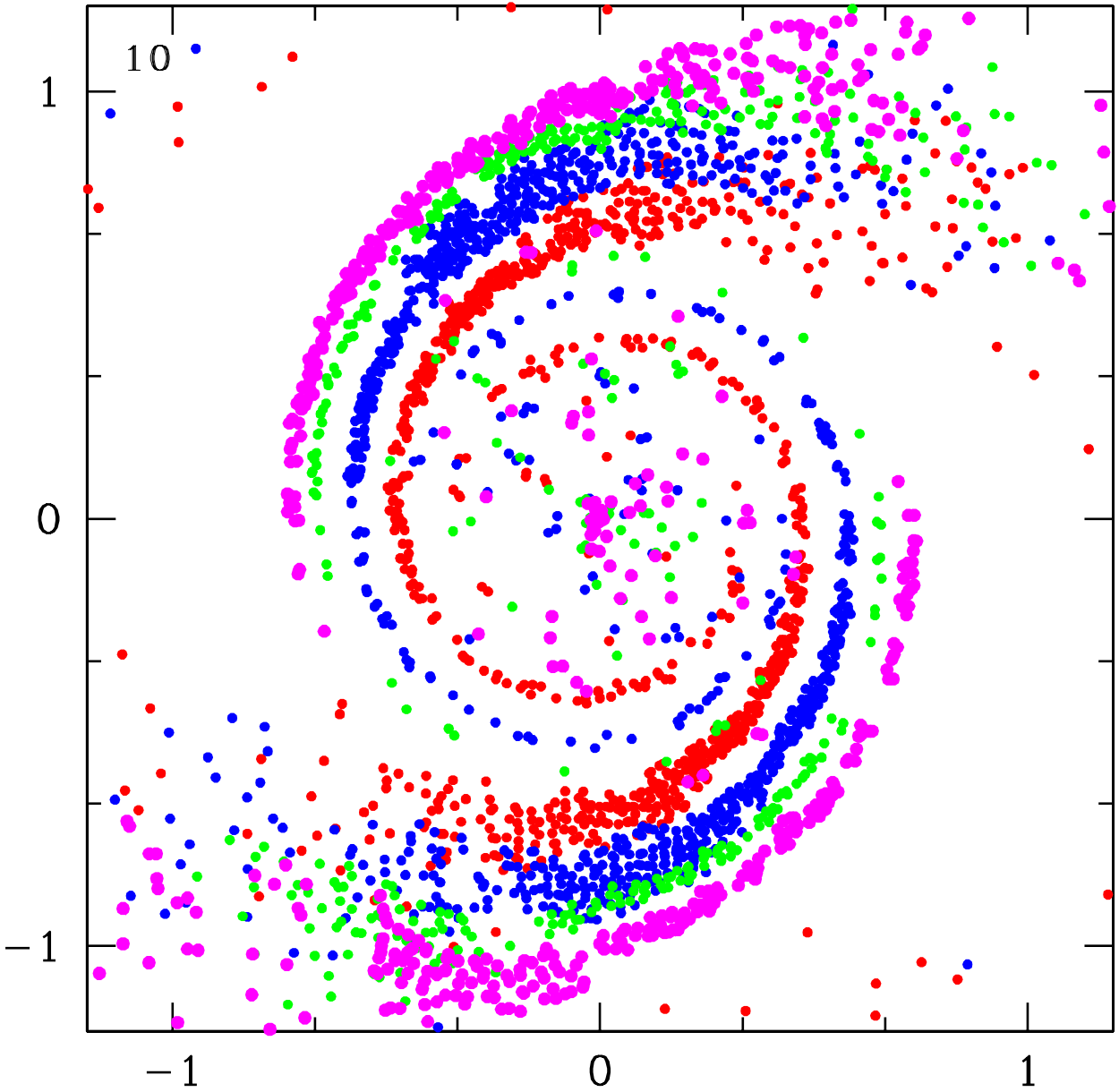}
\figcaption{Nuclear distribution of the gas particles that are shocked
by supersonic flow (i.e., $-h {\bf \nabla} \cdot {\bf  \upsilon} > c_{s}$)
for models M1-M4  at the evolution time of $10 \tau_{bar}$. 
Red: model M1 with $\gamma=0$; blue: model M2 with $\gamma=0.5$; 
green: model M3 with $\gamma=1$; magenta:  model M4 with $\gamma=1.5$. 
The gas sound speed is $c_{s}=10$ km s$^{-1}$. The number in the top left 
corner of the panel is the evolution time in units of $\tau_{bar}$. 
The bar lies horizontally, and the box size is $2.4$ kpc in one dimension.}
\end{figure}
\clearpage
\begin{figure}
\plotone{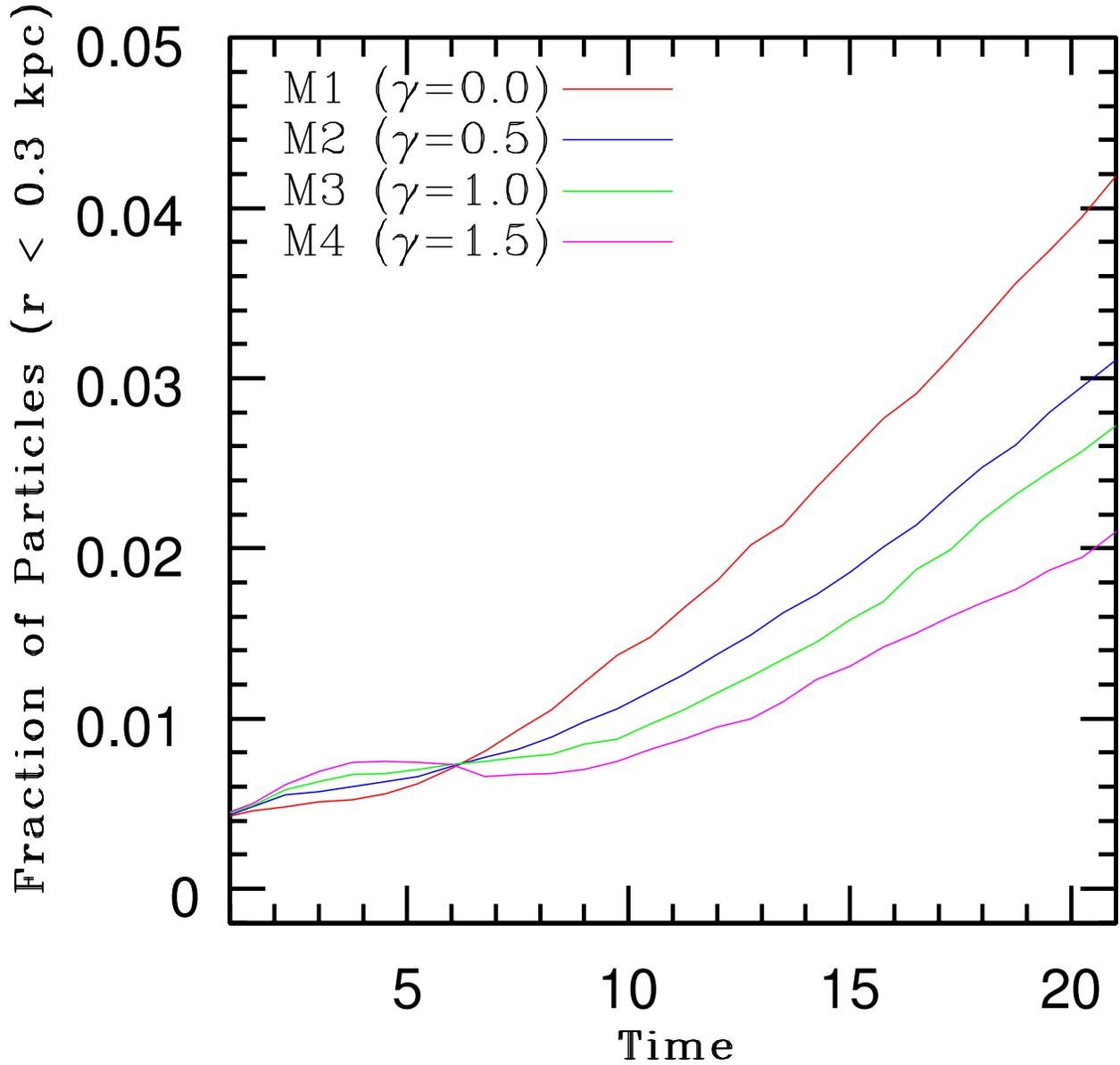}
\caption{The inflow of the gas particle into the central hundred parsec 
of models M1-M4. The time evolution of the fraction
of gas particles accumulated inside a radius of $0.3$ kpc is
shown. The time is given in units of $\tau_{bar}$. 
The gas sound speed is $c_{s}=10$ km s$^{-1}$. The line colors
corresponding to each model are given in the top left corner
of the panel.}
\end{figure}
\clearpage
\begin{figure}
\plotone{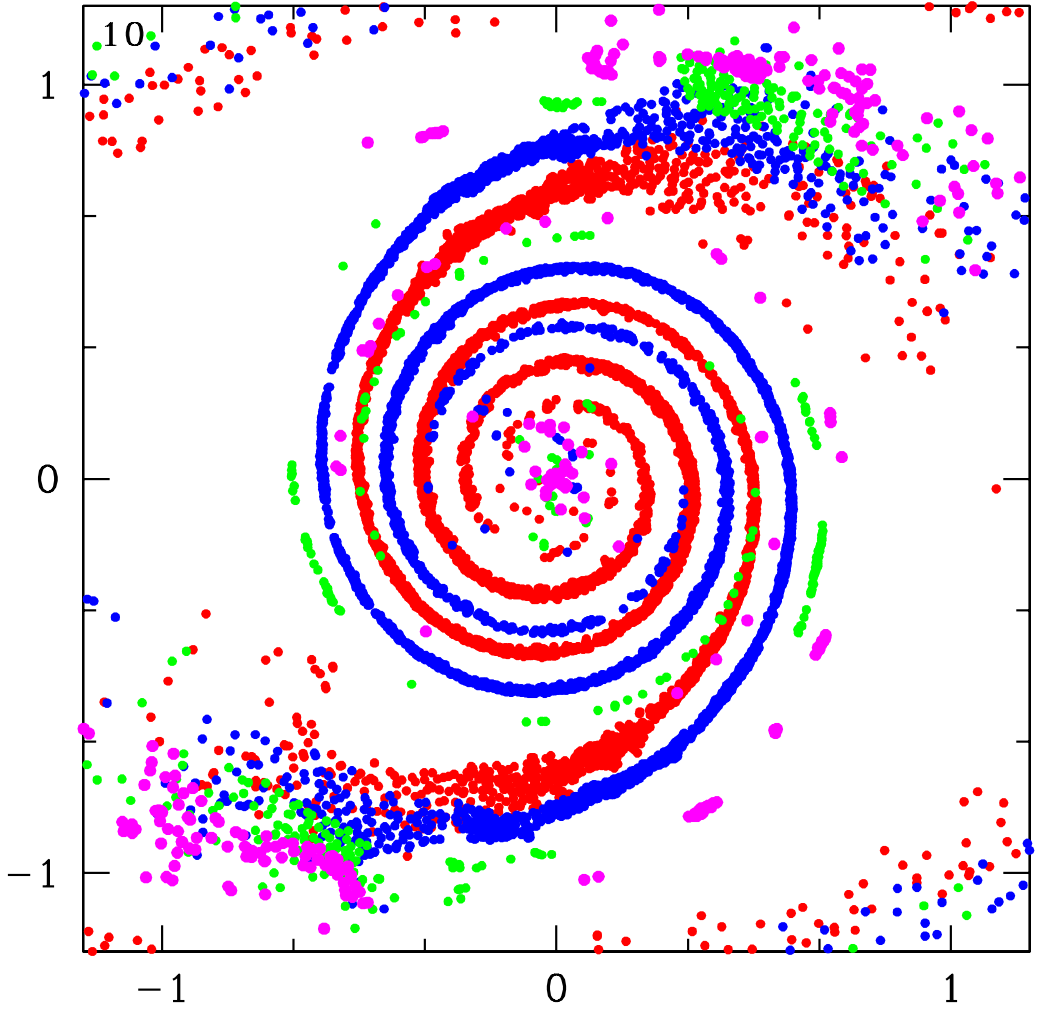}
\caption{Same as Fig. $3$, but for $1 \times 10^6$ SPH particles.}
\end{figure}

\begin{thebibliography}{}
\bibitem[Ann(2001)]{ann01} Ann, H.B. 2001, \aj, 121, 2515
\bibitem[Ann \& Lee(2004)]{ann04} Ann, H.B., \& Lee, H.M. 2004,
\apj, 613, L105
\bibitem[Ann \& Thakur(2005)]{at05} Ann, H.B., \& Thakur, P. 2005,
 \apj, 620, 197
\bibitem[Athanassoula(1992)]{ath92} Athanassoula, E. 1992,
    \mnras, 259, 345
\bibitem[Balsara's(1995)]{bal95} Balsara, D.S. 1995, J. Comput.
    Phys., 121, 357
\bibitem[Benz(1990)]{ben90} Benz, W. 1990, in The Numerical
    Modelling of Nonlinear Stellar Pulsations, ed. J.R. Buchler
    (NATO ASI Ser. C 302; Dordrecht: Kluwer Academic), 269
\bibitem[Block et al.(2001)]{blk01} Block, D.L., Puerari, I., Knapen, J.H., 
        Elmegreen, B.G., Buta, R., Stedman, S., \& Elmegreen, D.M. 2001, \aap, 375, 761
\bibitem[Buta \& Combes(1996)]{buta96} Buta, R.J., \& Combes, F. 1996, 
     Fund. Cosmic Phys., 17, 95
\bibitem[Carollo et al.(2002)]{car02} Carollo, C.M., Stiavalli, M., 
Seigar, M., de Zeeuw, P.T., \& Dejonghe, H. 2002, \aj, 123, 159 
\bibitem[Combes et al.(2002)]{com02} Combes, F., Boisse, P., Mazure, A.,
   \& Blanchard, A. 2002, Galaxies and Cosmology (New York: Springer), 172
\bibitem[Cowei(1980)]{cow80} Cowei, L.L. 1980, \apj, 236, 868
\bibitem[Dehnen(1993)]{deh93} Dehnen, W. 1993, \mnras, 265, 250 
\bibitem[Elmegreen et al.(1998)]{elm98a} Elmegreen, D.M., Chromey, F.R.,
\& Warren, A.R. 1998, \aj, 116, 2834
\bibitem[Elmegreen \& Thomasson(1993)]{elm93} Elmegreen, B.G., 
    \& Thomasson, M. 1993, \aap, 272, 37
\bibitem[Englmaier \& Gerhard(1997)]{eng97} Englmaier, P.,
    \& Gerhard, O. 1997, \mnras, 287, 57
\bibitem[Englmaier \& Shlosman(2000)]{eng00} Englmaier, P.,
    \& Shlosman, R.S. 2000, \apj, 528, 677
\bibitem[Freeman(1966)]{fre66} Freeman, K.C. 1966, \mnras,
    134, 1
\bibitem[Freeman's(1970)]{fre70} Freeman, K.C. 1970, \apj,
    160, 811
\bibitem[Fukuda et al.(2000)]{fuk00} Fukuda, H., Habe, A.,
    \& Wada, K. 2000, \apj, 529, 109
\bibitem[Fukuda et al.(1998)]{fuk98} Fukuda, H., Wada, K.,
    \& Habe, A. 1998, \mnras, 295, 463
\bibitem[Fux(1999)]{fux99} Fux, R. 1999, \aap, 345, 787
\bibitem[Fux(2001)]{fux01} Fux, R. 2001, JKAS, 34, 255
\bibitem[Hasan et al.(1993)]{has93} Hasan, H., Pfenniger, D., \& Norman, C.
1993, \apj, 409, 91
\bibitem[Hut et al.(1995)]{hut95} Hut, P., Makino, J.,
    \& McMillan, S. 1995, \apj, 443, L93
\bibitem[Kamphuis(1993)]{kam93} Kamphuis, J. 1993, Ph.D. Thesis, Univ. Groningen
\bibitem[Knapen et al.(1995)]{knp95} Knapen, J.H., Beckman, J.E.,
Heller, C.H., Shlosman, I., \& de Jong, R.S. 1995, \apj, 454, 623
\bibitem[Knapen(1999)]{knp99} Knapen, J.H. 1999, in the Evolution of
Galaxies on Cosmological Timescales, ed. J.E. Beckman, \& T.J. Mahoney
(ASP Conf. Ser. 187), 72
\bibitem[Laine et al.(1999)]{lai99} Laine, S., Knapen, J.H.,
    Perez-Ramiez, D.,
 Doyon, R., \& Nadeau, D. 1999, \mnras, 302, L33
\bibitem[Laine et al.(2001)]{lai01} Laine, S., Knapen, J.H.,
    Perez-Ramiez, D.,
 Englmaier, P., \& Matthias, M. 2001, \mnras, 324, 891
\bibitem[Lee et al.(1999)]{lee99} Lee, C.W., Lee, H.M., Ann, H.B., 
    \& kwon, K.H. 1999, \apj, 513, 242
\bibitem[Long \& Murali(1992)]{lon92} Long, K., \& Murali,
    C. 1992, \apj, 397, 44
\bibitem[Maciejewski et al.(2002)]{mac02} Maciejewski, W.,
    Teuben, P.J., Sparke, L.S., \& Stone, J.M. 2002, \mnras, 329, 502
\bibitem[Maciejewski(2003)]{mac03} Maciejewski, W. 2003,
    in the Proc. JENAM 2002, Galactic and Stellar Dynamics, ed. C. Boily et al.
 (Les Ulis: EDP Sciences), 3
\bibitem[Maciejewski(2004)]{mac04} Maciejewski, W. 2004, \mnras, 354, 892
\bibitem[Martini \& Pogge(1999)]{mar99} Martini, P.,\&
    Pogge R.W. 1999, \aj, 118, 2646
\bibitem[Martini et al.(2003a)]{mar03a} Martini, P., Regan, M.W.,
    Mulchaey, J.S., \& Pogge, R.W. 2003a, \apjs, 146, 353
\bibitem[Martini et al.(2003b)]{mar03b} Martini, P., Regan, M.W.,
    Mulchaey, J.S., \& Pogge, R.W. 2003b, \apj, 589, 774
\bibitem[Monaghan(1992)]{mon92} Monaghan, J. 1992, \araa, 30, 543
\bibitem[Namekata et al.(2008)]{name08} Namekata, D., Habe A., Matsui, H., 
\& Saitoh, T.R. 2008, \apj, Submitted (arXiv:0810.1095v1)
\bibitem[Patsis \& Athanassoula(2000)]{pat00} Patsis, P.A.,
    \& Athanassoula, E. 2000, \aap, 358, 45
\bibitem[Perez(2008)]{per08} Perez, I. 2008, \aap, 478, 717
\bibitem[Pfenniger \& Friedli(1993)]{pfe93} Pfenniger, D.,
    \& Friedli, D. 1993, \aap, 270, 561
\bibitem[Phillips et al.(1996)]{phi96} Phillips A.C., Illingworth,
    G.D., MacKenty, J.W., \& Franx, M. 1996, \aj, 111, 1566
\bibitem[Piner et al.(1995)]{pin95} Piner, B.G., Stone, J.M., \& Teuben, P.J.
1995, \apj, 449, 508 
\bibitem[Pogge \& Martini(2002)]{pog02} Pogge R.W., \&  Martini,
    P. 2002, \apj, 569, 624
\bibitem[Regan \& Mulchaey(1999)]{reg99} Regan M.W.,
    \& Mulchaey J.S., 1999, \aj, 117, 2676
\bibitem[Selloow \& Wilkinson(1993)]{sell93} Selloow, J.A., \& Wilkinson, A. 1993, 
Rep. Prog. Phys., 56, 173
\bibitem[Steinmetz(1996)]{ste96} Steinmetz, M. 1996, \mnras, 278, 1005
\bibitem[Wada(1994)]{wad94}  Wada, K. 1994, \pasj, 46, 165

\end{thebibliography}
\end{document}